\documentstyle [12pt,a4]{article}
\newcommand{\mincir}{\raise -2.truept\hbox{\rlap{\hbox{$\sim$}}\raise5.truept
\hbox{$<$}\ }}
\newcommand{\magcir}{\raise -2.truept\hbox{\rlap{\hbox{$\sim$}}\raise5.truept
\hbox{$>$}\ }}
\newcommand{\minmag}{\raise-2.truept\hbox{\rlap{\hbox{$<$}}\raise 6.truept\hbox
{$>$}\ }}
\newcommand{\be}{\begin{equation}}
\newcommand{\ee}{\end{equation}}
\newcommand{\ba}{\begin{eqnarray}}
\newcommand{\ea}{\end{eqnarray}}
\newcommand{\brr}{\begin{array}}
\newcommand{\err}{\end{array}}
\newcommand{\bc}{\begin{center}}
\newcommand{\ec}{\end{center}}

\newcommand{\bx}{\mbox{\bf x}}

\newcommand{\lb}{{\left<\right.}}
\newcommand{\rb}{{\left.\right>}}
\newcommand{\hm}{\,h^{-1}{\rm Mpc}}

\title{ {\bf The Angular Distribution of Clusters in Skewed CDM Models}}
\author{
{\bf Stefano Borgani}$^1$, {\bf Peter Coles}$^2$, \\
{\bf Lauro Moscardini}$^3$ and {\bf Manolis Plionis}$^4$ \\ ~\\
{\em $^1$INFN, Sezione di Perugia,}\\
{\em c/o Dipartimento di Fisica dell'Universit\`a,} \\
{\em via A. Pascoli, I--06100 Perugia, Italy} \\ ~\\
{\em $^2$Astronomy Unit, School of Mathematical Sciences,} \\
{\em Queen Mary \& Westfield College, Mile End Road,} \\
{\em London, E1 4NS, UK} \\ ~\\
{\em $^3$Dipartimento di Astronomia, Universit\`a di Padova,} \\
{\em vicolo dell'Osservatorio 5, I--35122 Padova, Italy} \\ ~\\
{\em $^4$SISSA -- International School for Advanced Studies,} \\
{\em via Beirut 2--4, I--34014 Trieste, Italy}  }

\date{}

\begin{document}

\maketitle
\vspace{0.5truecm}
\centerline{\it Ref. DFUPG 74--93}
\vspace{0.3truecm}
\centerline{\sl The Astrophysical Journal Letters, submitted}
\thispagestyle{empty}

\setcounter{page}{0}
\newpage
\section*{\center Abstract}
We perform a detailed investigation of the statistical properties of the
projected (angular) distribution of galaxy clusters obtained in Cold Dark
Matter (CDM) models with both Gaussian and skewed (i.e. non--Gaussian)
primordial density fluctuations. We use large numerical simulations of
these skewed CDM models to construct a set of simulations of the Lick
catalogue. An objective cluster--finding algorithm is used to identify
regions where the projected number--density of galaxies in the catalogues
exceeds some density threshold criterion. In this way we can construct
catalogues containing the angular position and richness of real and
simulated clusters which are suitable for statistical analysis. For
Gaussian models, the overall number of clusters is too small in the
standard CDM case compared to observations, but a model with higher
normalisation is in much better agreement; non--Gaussian models with
negative initial skewness also fit the observed numbers fairly well. We
compute the angular correlation function of clusters of different richness
and find a strong dependence of the clustering amplitude with richness in
all models. Even with a higher normalization, the Gaussian CDM model fails
at producing sufficient large--scale cluster clustering. We also find that
the Lick data are better reproduced only by a CDM model with negative
initial skewness; initially skew--positive models fail to produce enough
large--scale clustering. This conclusion is confirmed by two other
statistical analyses; the properties of the minimal spanning tree and the
multifractal scaling of the real clusters are much better reproduced by
skew--negative CDM models. In particular, the small--scale self--similarity
in the distribution of richest real clusters turns out to be a crucial
test, which is only passed by skew--negative models. We show that a
skewness--variance relation of hierarchical type is followed by
skew-positive models, as well as by the more evolved Gaussian model.

\vspace{0.5cm}
\noindent{\bf Key Words:} Galaxies: formation, clustering -- large-scale
structure of the Universe -- early Universe -- dark matter.
\vspace{0.5cm}

\newpage

\section{Introduction}
Clusters of galaxies have been recognised for some time to be efficient
tracers  of the large--scale structure of the Universe, especially on scales
$>10\hm$ where the clustering of galaxies themselves is hardly detectable
above the noise (see, e.g., Bahcall 1988). Rich galaxy systems are
strongly correlated even on scales where the gravitational dynamics of the
underlying matter distribution is still in the linear regime, so that their
spatial distribution should give useful information about the spectrum and
probability distribution function of the primordial density fluctuations,
which are thought to be the seeds of galaxy formation.

A classical result in the analysis of cluster clustering is
the power--law shape of the 2--point correlation function of clusters,
\be
\xi(r)~=~\left({r_o\over r}\right)^\gamma\,,
\label{eq:xi}
\ee
which declines with an exponent $\gamma \simeq 1.8$, remarkably similar to
that relevant for galaxies, but with a much larger correlation length,
$r_o\simeq 20\hm$ (Bahcall \& Soneira 1983; Klypin \& Kopylov 1983; Postman,
Huchra \& Geller 1992). Well--defined relations have also been
found between cluster richness and correlation amplitude, the richest clusters
being much more strongly correlated than relatively poor ones
(Postman, Geller \& Huchra 1986; Bahcall \& Burgett 1986). It is possible
that this is part of an even wider relation which involves single
galaxies and perhaps even huge superclusters. At least
qualitatively, this behaviour can be
interpreted in hierarchical clustering theories, such as
CDM, as a consequence of the fact that larger structures arise from
higher peaks in the primordial density fluctuations which are
intrinsically more strongly clustered than typical points
(Kaiser 1984). But even allowing for this effect,
the large observed amplitude of the cluster correlation
is difficult to account for quantitatively in CDM models. This lack
of large--scale power represented one of the first
pitfalls of the `standard' CDM model, which
predicts a correlation length smaller than the observed value
by a factor $\sim 2$ (White et al. 1987). Some authors have taken this line of
argument to an extreme and argued that the observed richness--clustering
relation is the consequence of a kind of fractal clustering, extending
up to arbitrarily large mass scales with no evidence of homogeneity
(Coleman \& Pietronero 1992). Independent analyses of the scaling
properties of the two-- and three--dimensional cluster distributions
(Borgani, Plionis \& Valdarnini 1993; Borgani, Mart\'{\i}nez \& Valdarnini
1993,
in preparation), together with the convergence of the cluster dipole
anisotropy at $\sim 150 \hm$ (Plionis \& Valdarnini 1991; Scaramella,
Vettolani \& Zamorani 1991), show that this explanation is not correct.

Serious doubts about the reality of the strong spatial correlations of
clusters have been raised by some authors (Sutherland 1988; Sutherland \&
Efstathiou 1991). They claim that the inclusion of background and/or
foreground galaxies can spuriously amplify the small--scale clustering in
the line--of--sight direction for a richness--limited sample such as the Abell
catalogue (Abell 1958). This claim is, however, still controversial.
Different authors have pointed out that, although present, this effect
should not be significant (Dekel et al. 1989; Olivier et al. 1990; Jing,
Plionis \& Valdarnini 1992). The availability of more recent cluster
samples based on automated procedures using plate scanning devices such as
the APM (Dalton et al. 1992) and COSMOS (Collins, Nichol \& Lumsden 1992),
should help
to clarify this point. Nevertheless, different groups still reach different
conclusions using these new samples, even though they consistently give a
correlation length in the range 13--$16\hm$. Efstathiou et al. (1992)
conclude that APM clusters display a quite low correlation length, similar
to that of the Abell clusters, after correcting the clustering anisotropy
in the line--of--sight direction. By contrast, Bahcall \& West (1992) observe
that APM clusters are generally poorer than Abell's, so that the weaker
clustering can be interpreted in terms of a general relation between
richness and clustering strength (Bahcall \& Burgett 1986; Postman, Geller
\& Huchra 1986). The large--scale coherence of the cluster distribution has
also been confirmed by recent analyses of the Postman, Huchra \& Geller
(1992) redshift sample. It should be stressed, however, that even allowing
for a correlation length as small as $13\hm$, both analytical arguments
based on linear theory (Coles 1988; Borgani 1990) and $N$--body simulations
(White et al. 1987) indicate that cluster clustering is still a problem for
the standard CDM scenario. More recently, Scaramella (1992),
Peacock \& West (1992) and Jing
\& Valdarnini (1993) have found that the power--spectrum traced by clusters
possesses much more large--scale power than the power--spectrum predicted in
the CDM model. The problem may even be more deeply rooted than the choice
of primordial fluctuation spectrum: Plionis, Valdarnini \& Jing (1992) have
discovered features in the spatial cluster distribution which are not
accounted for by Gaussian models constructed to reproduce the observed 2--
and 3--point cluster correlation functions.

In a series of papers, we have performed a detailed analysis of
the statistical properties of angular cluster samples, obtained from
the Lick map by Plionis, Barrow \& Frenk (1991; hereafter PBF samples).
These clusters are selected by the application of
an objective overdensity criterion to the underlying galaxy
cell--counts provided by the Lick map (see Section 3, below).
Plionis, Barrow \& Frenk (1991; Paper I)
analysed the projected shapes of clusters.
Plionis \& Borgani (1991; Paper II) and Borgani, Jing \& Plionis (1992;
Paper III) worked out the correlation properties of the PBF cluster
distributions, devoting particular care to the relation between cluster
richness and clustering strength. Borgani, Plionis \& Valdarnini (1993;
Paper IV) performed a detailed multifractal analysis of the PBF samples,
finding that, at least for the richest clusters, a self--similar clustering
develops at the scales of non--linearity (i.e., ~$\mincir 20\hm$).

In this paper, we apply a number of statistical analyses to angular
cluster samples obtained from $N$--body simulations with skewed CDM
initial conditions (Moscardini et al. 1991; Matarrese et al. 1991;
Messina et al. 1992; Lucchin et al.
1993). Cluster samples are selected by applying the same criteria as in
Paper I to the simulated Lick maps (Coles et al. 1993b).

It is important to consider the effect of skewed CDM initial conditions for
at least two main reasons. First, as already observed, the large--scale
clustering of rich galaxy systems represents one of the drawbacks of the
standard CDM model.  It is interesting to ask whether dropping the
assumption of initially Gaussian density fluctuations enables one to add
sufficient large--scale cluster correlation to reconcile the CDM model with
observational data. Indeed, previous statistical analyses of the
simulations of the models considered in this work have shown that models
with negative initial
skewness are at least as good as Gaussian models at reproducing the
observed clustering data (Moscardini et al. 1991; Messina et
al. 1992; Lucchin et al. 1993; Coles et al. 1993a).
In particular they succeed in producing coherence on large scales and
high bulk motions by the slow non--linear process of merging of voids and
disruption of low--density bridges. The 2--point correlation function and
the topological analysis applied to projected catalogues extracted from
these models give support to the same result (Moscardini et al. 1993;
Coles et al. 1993b).
Moreover Matarrese et al. (1991) showed that grouping properties are
strongly dependent on the statistics of the underlying mass distribution.
Second, the observed scale invariance of cluster distributions on scales
$R~\mincir 20\hm$ (Borgani, Plionis \& Valdarnini 1993)
bears all the hallmarks of strongly
non--Gaussian statistics. The question is whether such non--Gaussian
statistics can be generated on a scale where the gravitational evolution of
the matter distribution is still linear, just by the cluster selection
procedure. If this proves to be impossible we need to consider another
source for this non--Gaussian signature.

The plan of this paper is as follows. In Section 2 we briefly describe the
considered non--Gaussian CDM models and the procedure for extracting the
Lick samples
from the $N$--body simulations. In Section 3 we review the PBF cluster
identification algorithm and comment on some simple  properties of the
resulting simulated cluster catalogues. Detailed quantitative statistical
analyses of the simulated cluster distributions are contained in Sections
4, 5, 6 and 7, where we also compare them with analogous results obtained
for the real cluster samples studied in Papers I--IV. The aim of this work
is to constrain our models for the statistics of the primordial CDM density
fluctuations. The main elements of our statistical analysis are as follows:
in Section 4 we evaluate the two--point cluster correlation function for the
different models and cluster richness; in Section 5 we investigate the
higher--order correlation statistics using a graph--theoretical construction
called the minimal--spanning--tree (MST); Section 6 is devoted to a fractal
analysis of the synthetic cluster samples; in Section 7 we apply the skewness
test to cell counts and analyze the skewness--variance relation. A detailed
discussion of the main results and a summary of our conclusions is
contained in Section 8.

\section{The Lick Simulations}
For this analysis we need to generate realistic simulations of the angular
spatial distribution of galaxies resulting from CDM models with skewed
(i.e. non--Gaussian) initial fluctuations. The first step in obtaining
simulated projected catalogues is to perform $N$--body computations of the
spatial distribution of galaxies in such scenarios (e.g., Messina et al. 1992).
We have generated such angular simulations already in the context of a study
of the topology of projected catalogues in non--Gaussian models by Coles et al.
(1993b). The details of our simulation techniques are described fully in
that paper, so we shall only give a brief outline here.

We consider the same types of non--Gaussian initial fluctuation statistics
as Moscardini et al. (1991), Messina et al. 1992,
Lucchin et al. (1993) and Coles et al.
(1993a,b), namely the Lognormal ($LN$) and Chi--squared of order unity
($\chi^{2}$). We construct simulations such that these distributions apply
to the peculiar gravitational potential $\Phi$, before the fluctuations are
modulated by the CDM transfer function. The distributions split into two
different types of model -- positive ($LN_p$ and $\chi^{2}_p$) and
negative ($LN_n$ and $\chi^{2}_n$) -- according to whether the linear mass
fluctuations have positive or negative skewness. The models are constructed
so that $\Phi$ has the CDM power--spectrum
\be
{\cal P}_\Phi(k) = {9 \over 4} {\cal P}_0 k^{-3} T^2(k),
\ee
where ${\cal P}_0 ~k$ is the primordial Zel'dovich spectrum of density
fluctuations and $T(k)$ is the CDM transfer function
(e.g. Davis et al. 1985)
\be
T(k) = [1 + 6.8 k + 72.0 k^{3/2} + 16.0 k^2 ]^{-1},
\ee
having considered a flat universe with Hubble constant $h=1/2$ in units of
$100$ km sec$^{-1}$ Mpc$^{-1}$. Standardizing the spectrum in such a way
allows a direct comparison with the standard, i.e. Gaussian, CDM
(hereafter {\em G}) model.

We used a particle--mesh code with $N_p = 128^3$ particles on $N_g=128^3$
grid--points [more details are given by Messina et al. (1992)].
Computations were performed at the CINECA Centre (Bologna) on a Cray
YMP/432 running under UNICOS. The box--size of our simulations is $L=260
{}~h^{-1}$ Mpc; each particle has mass $m = 4.7 \times 10^{12} {\rm
M}_\odot$. We evolve our models starting from the same amplitude up to the
`present time' $t_0$. We define $t_0$ as the time when the galaxy
two--point function is best fitted by the power--law
$\xi(r) \propto r^{-\gamma}$, with $\gamma=1.8$ in a suitable interval. To
obtain the {\em galaxies} in a given simulation we proceed as follows: we
smooth the initial density field with a Gaussian filter of radius
$1~h^{-1}$ Mpc and identify as galaxies the set of particles inside
excursion sets obtained by a threshold fixed in order to have a galaxy
number density equal to $3 \times 10^{-2}~h^3$ Mpc$^{-3}$, corresponding to
a total number of $\sim 530,000$ galaxies in the whole box (suitable for
the construction of our Lick look--a--like maps).

Different epochs will be parameterized by the {\em bias factor} $b$
defined, as usual, from the variance of linear mass--fluctuations
on a sharp--edged sphere of radius $R_8=8~h^{-1}$ Mpc, i.e.
\be
\sigma^2(R_8) = {{\cal P}_0 \over 2 \pi^2} \int_0^\infty dk k^3
T^2(k) W^2_{TH}(kR_8)  = {1\over b^2},
\ee
where $W_{TH}(x)=(3/x)j_1(x)$ is a top--hat window function and $j_1$ is
the Bessel function of order $1$. The present time $t_0$ corresponds to
$b=1.5$ for the Gaussian model, $b=2$ for both the positive models, $b=0.5$
for the negative $\chi^2$ and $b=0.4$ for the negative Lognormal. For this
study we use a subset of the simulations, comprising 6 in total. We use two
different times for the Gaussian simulation, $G_{1.5}$ and $G_1$, defined such
that $b=1.5$ (i.e. at the present epoch) and $b=1$ respectively. The four
skewed simulations, one of each type described above, are all considered at
the relevant present epoch.

The primordial gravitational potential, $\Phi(\bx )$ is obtained by the
convolution of a real function $\tau({\bf x})$ with a random field
$\varphi({\bf x})$,
\be
\Phi({\bf x}) = \int d^3 y \ \tau({\bf y} - {\bf x}) \varphi({\bf y}).
\ee
The field $\varphi$ is obtained by a non--linear transformation on
a zero--mean Gaussian process $w$, with unit variance and flicker--noise
power--spectrum; the function $\tau$ is fixed by its
Fourier transform,
\be
\tilde{\tau} ({\bf k}) \equiv \int d^3 x e^{-i {\bf k}
\cdot {\bf x}} \tau({\bf x}) = T(k) F(k),
\ee
where $T(k)$ is the CDM transfer function of eq.(3) and $F(k)$ a
positive correction factor which we applied to have the exact CDM initial
power--spectrum of eq.(2) in all our models. The precise forms
of the non--linear transformation from $w$ to $\varphi$ are
\be
\varphi({\bf x}) \propto e^{w({\bf x})},
\ee
and
\be
\varphi({\bf x}) \propto w^2({\bf x})
\ee
for $LN$ and $\chi^2$  respectively (Moscardini et al. 1991).

The Lick map has a characteristic depth of $D_{*} \sim 210$ $h^{-1}$ Mpc
(Groth \& Peebles 1977), but galaxies with $D \gg D_{*}$ are
also included in the
catalogue. In order to simulate the overall extent of the Lick
catalogue we therefore need to replicate our original simulation
box exploiting its periodic boundary conditions.
As the box--side is 260 $h^{-1}$ Mpc and the solid angle we want to study is
such that $b^{\mbox{\tiny II}} \ge 45^{\circ}$, we have to consider
the superposition, in the $z$--direction, of three levels of replicated boxes
while each level, starting from $z=0$, has an increasing number of boxes;
4 at the lowest level, 16 at the intermediate and 36 at the highest
level (i.e. 56 boxes).
Defining the $z$--cartesian axis in the direction of the Galactic north pole,
we choose the origin of our coordinate system to be at $z=0$ and at the central
point in the $x$, $y$ coordinates of the lowest series of connected boxes.

In order to generate our Lick look--a--like maps we assign to the
$\sim 530,000$ galaxies an absolute magnitude  according to the Schechter
(1976) luminosity function and then determine its apparent magnitude
taking into account $K$--corrections and expansion effects. We then
select galaxies whose apparent magnitude exceeds the corresponding value
for the Lick map ($m_{lim} \le 18.8$). If the number of selected galaxies
is larger than that of the Lick catalogue (316,000 for
$b^{\mbox{\tiny II}} \ge 45^{\circ}$) we perform a sparse
sampling in the set of the included objects.
We have tested the robustness of our simulated catalogues to variations of the
luminosity function in Coles et al. (1993b).

A possible problem, deriving from the use of replications of the original box,
could be an artificial and periodic magnification of the structures present
in the box. Note, however, that all the galaxies, even those which are
selected more than once, are assigned a different absolute magnitude
and since we are interested in the projected distribution of
galaxies, most of the artificial periodicity will be probably
washed out. In Coles et al. (1993b), we performed quantitative tests
of any residual spurious superposition and found the effect to be very small.
Grey--scale plots of the resulting angular distributions are displayed
in Coles et al. (1993b). The models differ widely in visual appearance; none
appears visually to match the behaviour of the Lick map in detail, but the
`most similar' appears to be the $\chi^{2}$ (negative) model which
seems to reproduce qualitatively the observed bubbly appearance
of the Lick counts. This is confirmed by the topology study which
demonstrates that the $\chi^{2}_{n}$ model provides a reasonable fit
to the quantitative measures of pattern used in that study.

\section{The cluster samples}
We extract cluster samples from the synthetic Lick catalogues (described
in the previous section) using the same procedure adopted in
Paper I by Plionis, Barrow \& Frenk (1991). The cluster
finding algorithm is based on identifying clusters as high peaks of the
underlying galaxy cell counts, after smoothing on a suitable angular scale.

If $n_{ij}$ is the unsmoothed galaxy count in the 10$\times 10$ arcmin
cell, labelled by the indices $i,j$, the corresponding smoothed count is
\be
n^*_{ij}\,=\,\sum_{l=i-1}^{i+1}\sum_{k=j-1}^{j+1}w_{lk}\,n_{lk}\,.
\label{eq:smn}
\ee
Here, the weights $w_{lk}$ are assigned so that $w_{ij}=1/4,~w_{i\pm 1,j}=
w_{i,j\pm 1}=1/8,~w_{i\pm 1,j\pm 1}=1/16$ and their sum is unity in order
to preserve the total galaxy count. This procedure is roughly equivalent
to smoothing the projected galaxy distribution with
a Gaussian window on a 30 arcmin scale.

After applying this smoothing procedure, we identify those cells whose smoothed
count is larger than a fixed threshold value:
\be
n^*_{ij}\,\ge \,\kappa \bar n\,,
\label{eq:nth}
\ee
$\bar n$ being the average cell count, which is obviously preserved after
smoothing. Connected cells whose galaxy count satisfies eq.(\ref{eq:nth})
form a cluster.
We identify the cluster centre
with the centre of the member cell having the highest galaxy density. This
definition is different from that adopted in Paper I, where the cluster
centre is identified as the centre of mass. We expect
this difference to have no significant influence on the statistics
of the cluster distribution on scales exceeding the 30 arcmin smoothing
scale. Selected clusters are found to have a quite well--defined shape,
with a central peak surrounded by cells with decreasing
density (see, e.g., Figure 2 of Paper I). In fact, we checked the
effect of taking the two definitions of cluster centres for the real
samples and found no differences between the respective clustering
properties.

By choosing different values for the threshold parameter $\kappa$ in
eq.(\ref{eq:nth}), we construct samples of clusters having different
richness, the richer corresponding to higher $\kappa$ values. Following
Paper II, we take four different thresholds, corresponding to $\kappa
=1.8,2.5,3$ and 3.6 (C18, C25, C30 and C36 samples, respectively).
Clusters identified at higher thresholds are also included in lower
$\kappa$ samples. It is worth remembering that such a cluster
identification algorithm is objective and, thus, does not introduce biases
arising from a visual inspection of photographic plates, as it is believed
to happen for the Abell (1958) and Zwicky et al. (1961) samples.

Our simulations have been set up to mimic only the northern galactic
hemisphere. We shall therefore compare them only with the clusters
seen in the northern part of the actual Lick map. Furthermore, there
is a strong dependence of the projected cluster
number density upon the galactic latitude at low $b^{\mbox{\tiny II}}$ values
(see, e.g., Paper II). We shall therefore apply a cutoff in galactic
latitude, using both real and simulated clusters only for
$b^{\mbox{\tiny II}}>50^\circ$ clusters.

We apply the above procedure to the 6 Lick simulations described in the
previous section, so that we end up with 24 cluster samples, whose
statistical properties are compared to that of the 4 observational data
sets. As an example of the resulting cluster distribution, we plot in
Figure 1 the observed as well as the simulated C25 samples in quasi
equal--area coordinates
\ba
X & = & (b^{\mbox{\tiny II}}-90^\circ)\,\sin l^{\mbox{\tiny II}} \nonumber \\
Y & = & (b^{\mbox{\tiny II}}-90^\circ)\,\cos l^{\mbox{\tiny II}}\,.
\ea
As usual, $b^{\mbox{\tiny II}}$ and $l^{\mbox{\tiny II}}$ represent the
galactic latitude and longitude, respectively. Also for the real data we
plot only $b^{\mbox{\tiny II}}> 45^\circ$ clusters. Already from a visual
inspection, it is apparent the remarkable difference in the number of
selected clusters and in their clustering pattern between different models.
In particular, some models like $G_{1.5}$, $LN_p$ and $\chi^2_p$, produce
many fewer clusters than observed. In this sense, Figure 1$a$ contains the
`bad' models, i.e. those that fail to account for the zero--th order
statistics of cluster number density. Figure 1$b$ contains those models
($G_1,~\chi^2_n$ and $LN_n$) that produce a larger number of clusters. As
far as the texture of the projected cluster distribution is concerned, we
note that the $LN_n$ model seems to produce too large features (voids,
filaments and cluster condensations) that, even after projection to a $\sim
210\hm$ depth, involve angular scales comparable to the width of the
observational cone. Our Figure 1 is to be compared with Figure 2 of Coles
et al. (1993b), which reports the whole galaxy distributions of the Lick
map simulations. By comparing these two pictures, it is easy to recognize
that cluster identification through peak selection acts as an amplifier of
the underlying clustering features and makes more clear the existing
difference between model and real data sets. The following sections contain
a more quantitative description of the statistics of the cluster
distributions.

More information about the selected cluster samples are presented in
Figure 2 and Table I, where we report the number of clusters and of active
(selected) cells, as well as the cluster richness, for both observational
data and simulations and for each adopted $\kappa$ value. Following Paper
II, we take two different definitions of cluster richness: either the
smoothed number count inside the most populated cell ($R_1$), or the average
count between the cells belonging to each cluster ($R_2$). Since $R_1$
values are found to have a smaller spread within the same sample, we plot only
this in Figure 2, while both $R_1$ and $R_2$, together with the respective
{\em rms} values, are given in Table I.

The left column of Figure 2 is for the `bad' models, while the right
column is for the `good' models. We note that the same richness--threshold
relation holds more or less for all the models, so it cannot be used to
discriminate between them. This is quite easy to understand because,
according to our definition of richness, $R_1$ essentially depends upon the
threshold chosen
for cluster identification. On the other hand, $G_{1.5}$, $LN_p$
and $\chi^2_p$ generate too few very dense cells, i.e. fewer
clusters than observed. Although $G_1$, $LN_n$ and $\chi^2_n$ are
much better, there is a systematic tendency to produce a number
of clusters  which is smaller by a factor $\sim 2$ with respect to
observations. It seems therefore that none of these CDM models, either
Gaussian or non--Gaussian, succeeds in generating as many clusters
as observed. We do not believe that this means that no CDM model is capable
of generating the number of clusters seen in the Lick map. Instead, we
think that the behaviour in the simulations can be ascribed to the finite
resolution of our $N$--body simulations. Since the
parent simulation cubes have a size of $260\hm$ with $128^3$ grid points,
we are not able to resolve scales below 2--3$\hm$, which, at the
depth of the Lick map ($\sim 210\hm$), corresponds to an angular size of
$\sim 50$ arcmin. Thus, even after smoothing the cell counts over the 30
arcmin scale used in the clustering identification procedure, we expect there
still to be a residual numerical smoothing, whose effect is to
suppress the projected density fluctuations and so to suppress
the number of cells belonging to high--threshold excursion sets.

In order to check whether this is the correct explanation and to see the
size of the effect of finite numerical resolution, we also applied the same
cluster identification procedure after binning the projected galaxy
distribution in 20$\times 20$ arcmin cells. This amounts to take an
effective smoothing scale of 1$^\circ$, thus larger than (or at least
comparable to) the scale of numerical smoothing. The results for
$\kappa=1.8$ and 2.5 are summarized in Table II, where we report for real
data and for each model the number of selected clusters and of active
cells. In this case we get qualitatively the same trends in the results at
higher angular resolution, but a much better agreement with real data is
obtained. The $G_{1.5}$, $LN_p$ and $\chi^2_p$ models still produce too few
clusters, so that they are virtually ruled out already by this test.
Although better, the $LN_n$ model gives a quite high number of C25 clusters
and, vice versa, a quite low number of C18 clusters.  This is not
surprising, since the $LN_n$ model generates large coherent structures, so
that, as the threshold decreases, many clusters percolate to form larger
structures. Therefore, their total number increases less than for a model
producing more isolated structures. Even in this case, the best models are
represented by the $G_1$ and $\chi^2_n$ models, which produce an adequate
number of clusters. On the one hand, these results confirm those obtained
by selecting clusters from 10$\times$10 arcmin cell counts; on the other,
they warn us that we must pay attention to the limited numerical
resolution. Nevertheless, we still prefer to adopt the original cluster
identification from counts in 10$\times$10 arcmin cells, since we wish to
compare the results we obtain in this paper with the previously--published
results on observational data sets, which are based on it. Only in those
cases where the comparison between data and simulations becomes
particularly problematic we will more carefully investigate the effect of
decreasing the angular resolution.

Obviously the large number of simulated samples we use generates a large
amount of information. To keep the paper down to a manageable size, we
shall discuss only the main results and concentrate on showing figures for
the `good' models (i.e., $G_1$, $LN_n$ and
$\chi^2_n$), while results for the `bad' models will be summarized in tables.

\section{The 2--point correlation analysis}
The angular 2--point correlation function, $w(\vartheta)$, for the PBF
clusters has been investigated in Paper II. Here we will use the estimator
\be
w(\vartheta)~=~\lambda^2\,{DD(\vartheta)\over RR(\vartheta)}-1\,,
\label{eq:wes}
\ee
where $DD(\vartheta)$ and $RR(\vartheta)$ represent the number of cluster
pairs at separation $\vartheta$ in the real sample and in a random sample
having the same boundaries as the real one. The random sample contains a
number of points which is $\lambda$ times larger than the real one. We find
that $\lambda =5$ is enough to get stable results. (Note that in Paper II,
$RR(\vartheta)$ was evaluated by averaging over 20 different random
samples, all having the same number of points as the real data set. Since
the total number of pairs increases as $\lambda^2$, we expect that the
procedure adopted here amounts to take 25 random realizations.) A
comparison of the results presented in Figure 3 for real clusters with
those in Figure 3 of Paper II shows a complete agreement.

In Figures 3$a,b$ and $c$ we show $w(\vartheta)$ for the $G_1$, $\chi^2_n$
and $LN_n$ models, as compared to real data, at each $\kappa$ threshold
value. The error bars have been estimated through the bootstrap resampling
technique (Ling, Frenk \& Barrow 1986). Also plotted are the best
least--square fit models
\be
w(\vartheta)~=~A\,\vartheta^{1-\gamma}
\label{eq:wpl}
\ee
for both real clusters (dotted lines) and simulations (dashed lines). In
Paper II we fixed $\gamma =2$ so to leave the amplitude $A$ as the only
parameter to be fitted. Here we prefer to leave free both $A$ and $\gamma$,
since different models develop significantly different slopes. In Table
III we show the results for all the models considered, along with the
corresponding {\em rms} uncertainties and the angular scale range where
eq.(\ref{eq:wpl}) provides a satisfactory fit.

As for the $G_1$ model, we note that it gives a systematically smaller
correlation amplitude at scales $>1^\circ$, while providing
consistent results at small angles. We observed that, since the
simulated clusters are identified from smoother distributions than the real
ones, they correspond at a fixed $\kappa$ value to relatively higher, and
consequently more strongly clustered, peaks.
This makes even more significant the lack of
correlation displayed by the $G_1$ clusters, at any $\kappa$ value.
Therefore, we conclude that Gaussian CDM models are not able to account for
the large--scale power traced by the cluster distribution, even when
identifying the present dynamical time in $N$--body simulations with more
evolved (i.e., $b\simeq 1$) configurations.

In contrast, the $LN_n$ model always produces systematically higher
large--scale correlations, with a slope of $w(\vartheta)$ which is smaller than
observed (see Figure 3$c$). This can be easily interpreted by looking at the
distribution of $LN_n$ clusters, as shown in Figure 1: the presence of huge
coherent structures generates an excess of large--scale clustering and does
not allow $w(\vartheta)$ to decline rapidly. We also investigated the effect
of taking a less evolved $LN_n$ configuration. Although in this case a correct
small--scale clustering amplitude is recovered, large--scale
phase coherence due to the highly non--Gaussian primordial statistics
generates an even flatter profile of $w(\vartheta)$.

Better results are obtained from the $\chi^2_n$ model. In this case (see
Figure 3$b$) both the clustering amplitude and the slope are better
reproduced. The still existing discrepancy could well be due to the fact
that our simulated clusters correspond to relatively higher peaks than real
ones. As a consequence, we expect to have a higher $w(\vartheta)$, with a
slightly larger slope.

As far as the `bad' models are concerned, we see from Table III that they give
unacceptably steep profiles of $w(\vartheta)$, thus confirming that all of them
fail at accounting for the clustering in the distribution of galaxy clusters.

As already observed in Paper II, the increasing trend of the correlation
amplitude with the $\kappa$ threshold translates into a clustering--richness
dependence. In Figure 4 we plot this relation for both the `good' models
and the observational data. A strong correlation is always generated,
although the details depend on the model. In order to compare
data and simulations, we shall remember again that our simulated clusters
have for a given $\kappa$ (thus for a given richness) an excess clustering
amplitude. This could be the explanation for the discrepancy between the
$\chi^2_n$ and $LN_n$ models and real data, while it should make even less
reliable the $G_1$ model, which already produces too a low clustering
strength.

As a brief summary of the correlation analysis of the cluster samples, we
can say that the $G_{1.5}$, $LN_p$ and $\chi^2_p$ clearly fail to account for
observational results. Even within the `good' models, the $G_1$ one does
not produce enough large--scale clustering. Both the $\chi^2_n$ and the
$LN_n$ gives reasonable results, although the former seems to be better
suited to provide an adequate amount of large--scale correlation amplitude.

\section{The MST analysis}
The {\em minimal spanning tree} (MST) for a given point distribution is defined
as the unique graph connecting all the points, with no closed loops and
having minimal length (e.g., Ore 1962). The construction of the MST graph
proceeds as follows. For our angular samples, let us start with a
randomly chosen cluster and connect
it to its nearest neighbour. At this first step, the tree $T_1$ has only one
branch of length $\vartheta_1$. At the $k$--th step, we define the distance of
the $i$--th point, still not belonging to the MST, from the $T_{k-1}$ tree as
\be
\vartheta_{i,T_{k-1}}~=~\min_{j\in T_{k-1}}\,\vartheta_{ij}\,.
\label{eq:dmst}
\ee
Thus, for a distribution of $N$ points the MST is given by $T_{N-1}$ and
contains the set of branch lengths $\{\vartheta_i\}_{i=1}^{N-1}$. From its
definition, it follows that the MST construction is unique and independent
of the point which is used to start building the tree. There are several
reasons why the MST is a useful tool in clustering analysis. First of all,
it is completely determined only once the position of each single point is
known, so that it conveys informations about correlations of any order.
Moreover, when one branch is added to the tree, its position does not
depend on that of the previously added branch, so that we can say that the
MST construction is delocalized. For the above reasons, the MST is
particularly suited to emphasize the main features of the global texture of
a point distribution, such as its connectivity, filamentarity, etc. The MST
has been applied in a cosmological context by Barrow, Bhavsar \& Sonoda
(1985) who showed that it is efficient at discriminating between different
kinds of models. Bhavsar \& Ling (1988) used the MST to study the
filamentarity of the spatial galaxy distribution in the CfA redshift
survey, while Plionis, Valdarnini \& Jing (1992) performed a similar
analysis using a redshift sample of Abell and ACO clusters. The construction
of the MST has also been recognised as a useful tool for estimating the
fractal dimension of a point distribution (van de Weygaert, Jones \&
Mart\'{\i}nez 1992).

In Figure 5 we show the edge--length frequency distribution, $F(\vartheta)$,
for the $\kappa=1.8$ (left column) and $\kappa=3$ (right column) clusters
generated by the `good' models. The $F(\vartheta)$ distribution is defined
in such a way that $F(\vartheta)\,d\vartheta$ is the fraction of branches
in the tree with length between $\vartheta$ and $\vartheta +d\vartheta$.
Barrow, Bhavsar \& Sonoda (1985) have shown that a Poissonian distribution is
characterized by a Gaussian $F(\vartheta)$, with the maximum
occurring at the average branch length, $\lb \vartheta \rb$. In Figure 5,
the solid histograms are for real data, while the dashed ones are for the
simulations. By comparing the results for the two different thresholds, one
can see again the increasing clustering strength with $\kappa$ value,
which is reflected in the increasing skewness of the $F(\vartheta)$
profile. The spanning trees of the different models are all different, but
the really striking differences are between the model trees
and the tree of the real cluster distribution.
In particular, all the simulations have broader and
more skewed $F(\vartheta)$ distributions, with an excess of
both very short and very long branches. Once more, this is probably due to
the fact that the simulated clusters represent higher peaks in the
background galaxy distributions than the real clusters. In order to check the
consistency of the $F(\vartheta)$ distributions in a more quantitative
way, we applied a non--parametric Kolmogorov--Smirnov (KS) test to the
various distributions. In Table IV we summarize the
results for the `good' models. All the models are rejected at a quite high
confidence level, probably because the
simulated samples contain systematically higher peaks.

In order to avoid such problems, we repeated the MST analysis for the
C18 clusters, identified from the $1^\circ$ smoothed count (see \S3). The
resulting $F(\vartheta)$ shows a much better agreement with real data, although
even in this case the $\chi^2_n$ model fares marginally better than the
others. This is also confirmed by the results of the KS test, reported in
Table IV. This shows that the capability of the MST to enhance
high--order statistical information makes it a useful instrument for
detecting differences between models, even in presence of rather
limited data sets.

\section{A fractal analysis}
Statistical methods based on fractal analysis are particularly suited to
investigate the scaling properties of point distributions, in several
physical contexts (e.g., Mandelbrot 1982).
A simple concept which is useful in understanding scaling properties
is the fractal, or Hausdorff, dimension. For a
given point distribution in a $d$--dimensional ambient space ($d=2$ for our
angular cluster distributions), let $N_c (\epsilon)$ be the number of cells of
size $\epsilon$, which are required to completely cover the set. For a
fractal distribution, we expect that at very small $\epsilon$ values,
$N_c(\epsilon)$ should scale according to a pure power--law,
\be
N_c(\epsilon)~\propto~\epsilon^{-D_o}\,.
\label{eq:d0}
\ee
Here, $D_o$ is defined as the box--counting or capacity dimension, which, in
most practical cases, gives a close estimate of the Hausdorff dimension.
According to the definition (\ref{eq:d0}), it must be $0< D_o\le d$. Note
that the capacity dimension depends only on the geometry of the
distribution via the number of non--empty cells. It does not measure anything
to do with the clustering properties of the set as measured by the
correlation functions. To characterize this sort of information,
a continuous set of scaling indices
must be introduced, which in some sense are equivalent to the whole
sequence of correlation functions, required to describe any statistical
system. To this purpose, let us define $p_i(\epsilon)=n_i(\epsilon)/N$, as
the probability measure in the $i$--th cell, containing $n_i(\epsilon)$ out
of the total number $N$ of points. Accordingly, the generalized Renyi
dimensions (Renyi 1970) are defined as
\be
D_q~=~{1\over q-1}\,\lim_{\epsilon \to 0}
{\log{\sum_i[p_i(\epsilon)]^q}\over \log \epsilon}
{}~~~~~~~~;~~~~~~~D_1~=~\lim_{q\to 1}D_q\,.
\label{eq:dq}
\ee
Note that for $q=0$ the above equation reduces to the definition
(\ref{eq:d0}) of capacity dimension. The $q=2$ case is for the correlation
dimension, which, at small scales, is related to the slope of the 2--point
correlation function according to $D_2 =3-\gamma$. The {\em multifractal}
dimension spectrum of eq.(\ref{eq:dq}) gives a comprehensive description
of the scaling properties of a point distribution; positive--order
dimensions are sensitive to the statistics inside the overdense regions,
while the negative--$q$ ones account for the underdensities. It is also
possible to show that, under general conditions, $D_q$ must be a
non--increasing function of $q$. A particularly simple case occurs when
$D_q=const$, independent of $q$. In this case, the distribution is said to
be monofractal and a single scaling index completely specifies the
statistics. More in general, the shape of $D_q$ gives a clear indication of
the nature of the clustering (e.g., Jones, Coles \& Mart\'{\i}nez 1992) and
precise relations can be found between the hierarchy of correlation
functions and the multifractal dimension spectrum (Balian \& Schaeffer
1989; Borgani 1993). A detailed review of multifractal concepts is given by
Paladin \& Vulpiani (1987).

The formal definition (\ref{eq:dq}) of multifractal
dimensions is given in the limit of infinitesimally small scales. Obviously,
in practical cases only a finite number of points is available
and only a limited scale range can be studied. In addition, in most
physical situations, different scaling regimes are expected at different
scales, thus giving rise also to an upper limit to the scale range where
self--similarity develops. This is just what happens in the study of the
large--scale structure, where the distributions of galaxies and galaxy
systems develop a well defined scaling only at small scales, while
homogeneity is recovered at large scales.

Fractal analysis methods have been applied in recent years in cosmology to
describe the scaling properties of galaxy clustering, using both
observational data and $N$--body simulations. The emerging picture is that
any self-similar behaviour is confined to small scales ($<5\hm$), where
clustering is non--linear (e.g., Mart\'{\i}nez et al. 1990; Valdarnini, Borgani
\& Provenzale 1992; Colombi, Bouchet \& Schaeffer 1992; Yepes,
Dom\'{\i}nguez--Tenreiro \& Couchman 1992).

In Paper IV we realized a detailed
multifractal analysis of the PBF cluster samples by applying several
dimension estimators. We found that the C36 clusters develop a good
scaling behaviour up to angular scales $\sim 5^\circ$--$6^\circ$, which
correspond to a physical scale $\sim 20\hm$ at the depth of the Lick map.
It is clear that, although non--linear gravitational
clustering furnishes a dynamical mechanism for  generating  self--similarity,
it only does so up to scales of few Mpc. Thus, at the larger scales of cluster
clustering, the scaling behaviour detected cannot have a
dynamical origin. At such
scales, the non--linearity of the clustering has a statistical origin and
resides in the identification of clusters as exceptionally high peaks of
the underlying density field. The question thus arises as to whether the
scaling detected up to $\sim 20\hm$, which is the unique imprint of a
strong non--Gaussian statistics, is generated just by means of a high--peak
selection on a Gaussian background or requires something more. Here we
apply the same analysis as in Paper IV to our synthetic cluster samples in
order to check whether the existence of the scaling behaviour in the real
data constrains the models with primordial non--Gaussian fluctuations.

A number of different methods have been devised to provide estimates
of scaling dimensions from finite data sets; these rely
on different definitions of dimension and use different
kinds of approximation. One must be very careful to use appropriate
methods when dealing with poor statistics or when self--similarity develops
only over a limited scale--range (e.g., Borgani at al. 1993).

The first method we use
is based on the evaluation of the moments of counts of neighbours
(Grassberger \& Procaccia 1983; Paladin \& Vulpiani 1984). For our
angular cluster distributions, we estimate the partition function
\be
Z(q,\vartheta )~=~{1\over N}\,\sum_{i=1}^N[p_i(\vartheta)]^{q-1}\,,
\label{eq:zqr}
\ee
where $p_i(\vartheta)=n_i(\vartheta)/N$, with $n_i(\vartheta)$ the number of
neighbours within the angular distance $\vartheta$ from the $i$--th object.
For a fractal structure, we expect
\be
Z(q,\vartheta)~\propto~\vartheta^{\tau_q}\,,
\label{eq:zqr1}
\ee
with the resulting multifractal dimension spectrum given by
\be
D_q~=~{\tau_q \over q-1}\,.
\label{eq:dq1}
\ee
Though it converges rapidly for $q>1$, this method is extremely sensitive
to discreteness effects when $q$ is negative, and one must use a different
method to probe that regime.

A second method, which is better suited in the $q<1$ regime represents a
kind of inversion of the previous one; instead of counting the number of points
within a given radius, it is based on measuring the radius of the sphere
containing a fixed number of points (Grassberger, Badii \& Politi 1988). Let
$\vartheta_i(p)$ be the smallest radius of the disk centred at the $i$--th
point and containing $n=pN$ objects. Then, the partition function
\be
W(\tau,p)~=~{1\over N}\,\sum_{i=1}^N[\vartheta_i(p)]^{-\tau}
\label{eq:wp}
\ee
describes the scaling properties of the distribution. For a fractal structure,
$W(\tau,p)\propto p^{1-q_\tau}$, and the corresponding multifractal spectrum
is given by eq.(\ref{eq:dq1}).

A further characterization of a multifractal set can be given in terms of
the singularity spectrum (Hentschel \& Procaccia 1983). Let us define the
local dimension $\alpha_i$, such that the scaling inside the $i$--th cell is
$n_i(\vartheta)\propto \vartheta^{\alpha_i}$. Moreover, let $S(\alpha)$ be
the subset containing all the points around which the local dimension lies
in the range $[\alpha,\alpha+d\alpha]$. We define the {\em singularity
spectrum}, $f(\alpha)$, as the Hausdorff dimension of $S(\alpha)$. It is
possible to show that the two pairs of variables, $(q,\tau_q)$ and
$(\alpha,f(\alpha))$, are related by a Legendre transform:
\be
\alpha (q)~=~{d\tau_q\over dq}~~~~~~~~~~;~~~~~~~~\tau_q~=~q\alpha-f(\alpha)\,,
\label{eq:leg}
\ee
so that they give equivalent descriptions of a fractal structure. According
to eq.(\ref{eq:leg}), if a distribution is monofractal, then only one $\alpha$
value is allowed and the singularity spectrum degenerates into a single point,
$(D_o,f(D_o)=D_o)$. In general, multifractal behaviour is reflected
in a spread of $\alpha$ values within a finite interval, with its smallest and
largest values determined by the scaling inside the most overdense and
underdense regions, respectively:
\be
\alpha_{min}~=~\lim_{q\to +\infty}D_q~~~~~~~;~~~~~~~~
\alpha_{max}~=~\lim_{q\to -\infty}D_q\,.
\label{eq:alim}
\ee
In this sense, we can say that a multifractal structure is characterized
by local scaling properties (different local dimensions around different
points), while a monofractal structure has global scaling. Furthermore, the
decreasing slope of $D_q$ constrains $f(\alpha)$ to be a convex function,
with its maximum value equal to the Hausdorff dimension.

In Figure 7 and 8 we plot the $Z(q,\vartheta)$ partition function for the
C36 and C25 clusters, respectively.
(We have analysed the C30 and C18 samples with this method but we do not
show the results here; they represent intermediate cases.)
{}From left to right, we display the
results for the $G_1$, $\chi^2_n$ and $LN_n$ models, while upper and
lower figures are for $q=0$ and $q=4$, respectively. In each panel we plot
both $Z(q,\vartheta)$ and the corresponding local dimension,
$D_q(\vartheta)$, evaluated from a three point linear regression on the
partition function slope. For an ideal fractal structure, $D_q(\vartheta)$
should remain constant. In general, the $\theta$ range where it stays flat
gives a precise idea of the scales at which self--similarity (if any) takes
place. According to the definition (\ref{eq:zqr}), the amplitude of the
partition function is not normalized to be the same for two
distributions having the same scaling properties but a different number of
points. For this reason, we are only interested to compare the slopes of
$Z(q,r)$ for data and simulations and not their amplitudes; only the slopes
are related to the corresponding fractal dimensions. In order to correct
for boundary effects in the computation of $Z(q,\vartheta)$ we take as centres
only those clusters whose angular distance from the sample boundaries is
$<\vartheta$.

For $q=0$, which corresponds to the estimate of the Hausdorff dimension,
for both the C36 and the C25 samples the local dimension shows a smooth
transition from $D_o\simeq 0$ at small scales, to $D_q\simeq 2$ at $\vartheta
\sim 6^\circ$, which is the signature of large--scale homogeneity (at least
in the projected distribution). Note that all the three models generate a
homogeneous geometry of the distributions roughly at the same scale as
observational data. At smaller scales, the best model is the $LN_n$ one, which
correctly reproduces the partition function slope. The other two models
slightly underestimate $D_q(\vartheta)$, although $\chi^2_n$ seems to be
more successful than $G_1$, especially at small angles. This small--scale
behaviour can be interpreted by saying that the low number of objects does
not allow to resolve the underlying distribution, so that the algorithm
measures nothing but the dimension of each single point, which is indeed zero.

More interesting is the $q=4$ case for the C36 clusters. As already found
in Paper IV, the real cluster distribution displays a rather flat
$D_q(\vartheta)$ shape at $\vartheta ~\mincir 6^\circ$, which indicates
small--scale scaling. In contrast, the $G_1$ model, does not succeed in
generating any small--scale self--similarity. Instead, at $\vartheta ~\mincir
2^\circ$ the partition function rapidly flattens, so that the local
dimension declines to zero. Unlike the $q=0$ case, we do not believe that
this is only an effect of discreteness for two main reasons. First, for
$q>1$ eq.(\ref{eq:zqr}) assigns most weight to the overdense parts of the
distribution, where the sampling on the scales we are looking at should be
good. Second, if the limited statistics were the reason, we should expect
to find a similar behaviour even for the other two models, which produces a
comparable number of C36 clusters. However, this is not the case for both
the $\chi^2_n$ and $LN_n$ models. Instead, they better reproduces the
small--scale flat shape of the local dimension, with a resulting fractal
dimension $D_q~\mincir 1$ to characterize the clustering inside the
overdense regions. In particular, note the remarkable good agreement
between data and $LN_n$ model at $\vartheta ~\mincir 6^\circ$, although at
larger scales it generates too much clustering and, consequently, too small
a dimension.

As already observed in Paper IV, as lower thresholds are considered any
self--similarity disappears. This is confirmed by Figure 8, which is
the same as
Figure 7, but for the C25 clusters. For both data and simulations no
scale range exists where $D_q(\vartheta)$ remains nearly constant. Even in this
case, the $G_1$ cluster distribution again underestimates the dimensions
at $\vartheta ~\mincir 2^\circ$, while both $\chi^2_n$ and $LN_n$ fare much
better.

Although we have also realized a similar scaling analysis using the
$W(\tau,p)$ partition function, for reasons of space we prefer not to show the
partition function results here. We note, however,
that these results confirm those coming
from the $Z(q,\vartheta)$ function and extend them even into the
regime where $q<1$; only the
$\chi^2_n$ and $LN_n$ models develop a self--similar clustering up to
$\vartheta \sim 5^\circ$--$6^\circ$, similar to that detected for real data
(see Figures 2 and 10 in Paper IV for the results of the $W(\tau,p)$
analysis on the PBF samples).

In Figure 9 we plot the $D_\tau$ multifractal dimension spectrum for the two
non--Gaussian models and for real data. The dimensions have been evaluated by
a log--log linear regression  of the $W(\tau,p)$ values in the $p$ range
where self--similarity develops. Also plotted is the $f(\alpha)$
singularity spectrum defined by eq.(\ref{eq:leg}). Error bars are standard
deviations in the linear regression. No similar plot has been produced for
the $G_1$ model, since it does not produce any fractality in the cluster
distribution and no fractal dimension can be defined. Both the $\chi^2_n$
and the $LN_n$ models produce slightly lower dimensions. While this
difference is not significant for $\tau <0$, it is for $\tau>0$. This is
also reflected by the values taken by the local dimension $\alpha$; note
that the $\alpha_{min}$ value is always smaller for the non--Gaussian
simulations than for the real data, thus indicating the presence of
stronger singularities. This difference could be partly due to the fact
that the clusters selected in the simulations correspond to relatively
higher peaks than real clusters. In fact, in a multifractal structure, the
dimension of the distribution of high peaks is always smaller than that for
the distribution of lower peaks.

In brief, the main result of this section is that the self--similar
clustering displayed by the C36 cluster sample at $\vartheta
{}~\mincir 6^\circ$ (corresponding to $R~\mincir 20 \hm$ in the spatial
distribution) is not generated by the Gaussian model, instead it requires the
presence of initial phase correlations, such that provided by the
$\chi^2_n$ and $LN_n$ models.

\section{Cell--count Skewness}
The analysis of the skewness of cell counts has been widely advocated
in  recent times as a probe of the large--scale structure of
the galaxy distribution (Efstathiou et al. 1990; Saunders et al. 1991;
Loveday et al. 1992). Coles
\& Frenk (1991) described in detail the physical motivation for the usefulness
of skewness as a diagnostic for the distribution of cosmic structures.
In particular, because the skewness is the
lowest--order imprint of non--Gaussian statistics, it is likely to be
a powerful test of the nature of initial conditions.
Coles et al. (1993a) used the skewness of the three--dimensional
distribution to discriminate between different non--Gaussian models for
initial conditions; in this section we apply
the skewness analysis to our projected cluster distributions, for both real
data and `good' models.

Let us divide the surveyed area of the sky into $N_c$ cells of
side $\vartheta$ and
let $N_i$ be the object
number count within the $i$--th cell. Then, if
$\bar N=\sum_{i=1}^{N_c}{N_i}/N_c$ is the average count within such cells, we
define $\delta_i=N_i/\bar N-1$ to be the relative fluctuation in the number
count. The statistics of the distribution can be described in terms of the
moments
\be
\lb \delta^n\rb~=~{1\over N_c}\,\sum_{i=1}^{N_c}\delta_i^n\,.
\label{eq:deln}
\ee
For $n=2$, eq.(\ref{eq:deln}) defines
the variance, $\Sigma^2=\lb \delta^2\rb$; $n=3$ gives the skewness,
$\Gamma =\lb \delta^3\rb$. A serious limitation in the analysis of the
moments of cell counts occurs when small cells are considered, which
contain only a small number of objects. In cases where most cells
contain only one cluster, the distribution is said to be dominated
by shot--noise. In less extreme circumstances, the discrete nature of
the distribution always imposes a non--vanishing skewness.
It is common use to account for such discreteness effects by assuming that
the point distribution represents a Poissonian sampling of a continuous
density field. In this case, it can be shown that the variance $\sigma^2$
and the skewness $\gamma$ of the continuous density field are related to
those of the discrete realization  according to
\ba
\sigma^2 & = & \Sigma^2 -{1\over \bar N} \nonumber \\
\gamma   & = & \Gamma -{3\sigma^2 \over \bar N}-{1\over \bar N^2}
\label{eq:disc}
\ea
(Peebles 1980; Saunders et al. 1991; Coles \& Frenk 1991).
It must be stressed, however, that this is only a model of the discreteness
effects; the data need not be well represented by such a Poissonian sampling,
so the statistics obtained by subtracting off the discreteness terms in
eq. (\ref{eq:disc}) need not correspond to anything physical.
In our case we are dealing with clusters defined as peaks of the underlying
galaxy density field and this is far from a Poisson sample. The equations
(\ref{eq:disc}) are not expected to provide a valid discreteness correction
for our case. We have verified this suspicion with our data:
the `raw' cell--counts give very stable variance and skewness behaviour
while the `corrected' results display only noise. For
the remainder of this section we will not consider any kind of discreteness
correction.

Coles \& Frenk (1991) discussed the hierarchical
variance--skewness relation
\be
\gamma ~=~3\,Q\,\sigma^4\,.
\label{eq:hier}
\ee
An expression of this form is expected to hold in a variety of clustering
scenarios: hierarchical clustering in the non--linear regime; quasi--linear
growth of Gaussian perturbations; lognormal density distributions.
Even biased CDM models follow this form to some degree of accuracy. The exact
value of the $Q$ parameter depends on the details of the model; observational
results suggest $Q\simeq 1$ for galaxies and $Q\simeq 0.6$--0.8 for clusters.

In order to test the reliability of eq.(\ref{eq:hier}), we plot in Figure
10 $\log \Gamma$ vs. $\log \Sigma^2$ for the C30 and C18 cluster samples,
for both real and simulated data. Error bars correspond to one standard
deviation, as evaluated by the bootstrap resampling procedure.
The dashed lines correspond to the
relation (\ref{eq:hier}), with $Q=0.6$. The hierarchical model
provides a remarkably good fit in all the cases, independent of the
initial conditions and the richness of the clusters selected,
at least at the scales of non--linear
clustering. On the other hand, in the weak clustering regime
eq.(\ref{eq:hier}) no longer applies and the skewness rapidly declines.
The stability of the result is also confirmed by Table 5, where we report
the best--fit values for $Q$, along with the respective standard deviations.
Although Figure 10 gives the impression of a remarkably good fit,
the uncertainties in the $Q$ values are quite large. We do find,
however, that a non--vanishing skewness is always detected at a
$2\sigma$ level. The corresponding variance--skewness relation closely
follows eq.(\ref{eq:hier}), with $Q\simeq 0.6$.
Although this result demonstrates the
remarkable stability of the hierarchical model for the projected
cell--counts, it does show that the skewness is not effective
at distinguishing the details of the different models. In the case
studied in this paper, a combination of projection and limited
statistics acts to weaken the usefulness of this test compared to its
three--dimensional analogue.

\section{Discussion and Conclusions}
In this paper we have analysed the statistical clustering properties of
clusters obtained by applying an objective overdensity criterion to the
Lick map and various simulations of it. Analysing clusters in this way,
rather than the individual galaxies, is a very efficient way to pick out
the essential features of the underlying clustering pattern. The
calculation of clustering properties such as the two--point correlation
function for the entire mock Lick maps is of course possible and indeed has
already been done (Moscardini et al. 1993), but it is a
laborious task. Using only the high--density regions allows one to describe
the projected pattern at less cost in terms of work. Of course, some of the
clusters selected by our criterion may not represent true physical
associations of galaxies in three--dimensional space, but this does not
matter. Whatever they are, we compare objects selected in precisely the
same way in both the real data and the simulations so it is irrelevant to know
whether the clusters are bound structures or chance projections.

The clusters we analyse show up differences between the different models
even at the `zero'--order level. The number of clusters selected by
applying the same criterion to the different models depends very strongly
on the nature of the model and we can rule some of our models out with
extremely high confidence using just this simple statistic on its own. In
particular, the standard CDM model with Gaussian initial conditions and
$b=1.5$ generates a projected galaxy distribution which is too smooth and
consequently fails to account for the observed number of PBF clusters.
Although this finding seems to be at variance with respect to the findings
of White et al. (1987) about the number of Abell clusters produced in a
biased CDM model, nevertheless we should bear in mind that we are not
selecting precisely the same kind of objects. It has been suggested that
Gaussian CDM models might be satisfactory if  galaxy formation proceeds in
such a way that galaxies are not significantly more clustered than the dark
matter. In the `low--bias'  ($b\simeq 1$) CDM model (e.g., Couchman \&
Carlberg 1992) the present epoch is identified with a configuration which
is dynamically much more evolved that the standard, biased, version. The
greater dynamical evolution allows the model more time to build up
coherence on larger scales. Moreover, the recent detection of temperature
anisotropy in the cosmic microwave background (Smoot et al. 1992) seems to
require a low bias parameter if the temperature anisotropy is interpreted
as the imprint of primordial fluctuations in the dark matter density. We
have found that this model does indeed generate a larger number of clusters
such that, taking into account the numerical smoothing in our simulated
Lick maps, is in reasonable agreement with real data.

Although this model does survive the zeroth--order analysis, it runs into
trouble when we apply more detailed statistical descriptors. In particular,
the 2--point cluster correlation function has a much lower amplitude than
the observed correlation function and has a much steeper slope on large
scales. This is a consequence of the shape of the CDM primordial
fluctuation spectrum and is an unavoidable consequence of Gaussian models
which have no phase coherence capable of generating large--scale power on
scales where there is little primordial power.

Given the apparent failure of the Gaussian CDM models to account for all
the data, it is natural to ask whether the CDM hypothesis can be rescued by
the introduction of non--Gaussian primordial fluctuations. The initially
skew--positive CDM models we have considered fail in a similar way to the
Gaussian model in both number and correlation strength of clusters. These
models introduce a phase coherence only on small scales and cannot
alleviate the lack of large--scale power in the CDM spectrum. On the other
hand initially skew--negative models are generally successful at accounting
for the cluster number--richness function and the two--point correlation
length. This is particularly true for $\chi^{2}_n$ which is the most
successful of all the models we have considered.

Although the two--point correlation analysis obviously gives us
useful information about the nature of clustering in the models
and the real data, the visual texture, which is also strikingly different in
the models, is dominated by higher--order correlations which are
difficult to measure correctly. We therefore decided to subject
our models to three further statistical analyses, which are more sensitive
to high--order correlations and could therefore provide more effective
discrimination between the models and the data than the simpler statistics.

We used the Minimal Spanning Tree in an attempt to measure the intrinsic
`filamentariness' of the cluster distribution. The analysis shows up
nicely a number of quantitative differences between all the models and,
again, the $\chi^{2}_n$ model emerges as the one that agrees best with
the real data.

The multifractal spectrum of a data set reveals information about the scaling
behaviour of higher--order moments and is related closely to the non--Gaussian
character of the distribution (Balian \& Schaeffer 1989; Jones, Coles
\& Mart\'{\i}nez 1992; Borgani 1993). We again find that this kind of analysis
can distinguish between our models and the scaling behaviour we see,
particularly in the high density regions, clearly favours the
skew--negative models. A remarkable aspect of the multifractal analysis is
the wide difference between Gaussian and skew--negative models
in the produced scaling properties: for the C36 clusters, the Gaussian
model clearly fails to generate the self--similar behaviour displayed by real
data at the scales of non--linear clustering ($~\mincir 20\hm$; see Paper IV),
while both $\chi^2_n$ and $LN_n$ are successful. The sensitivity of this
test suggests that the observed self--similarity at scales where the
gravitational clustering is still in the linear regime can be justified on
the ground of initial phase correlations.

As a final test we also applied the skewness analysis, which represents the
lowest--order statistics to detect a non--Gaussian behaviour. We find that
a variance--skewness relation of hierarchical type is always satisfied, for
both the unbiased Gaussian model and the skew--negative models.

All these results are in good agreement with a previously--published
analysis of the same Lick map simulations using two--dimensional topological
characteristics by Coles et al. (1993b). We have also performed a
correlation analysis of individual galaxies in the simulated Lick maps,
comparing them with the results from the APM data (Moscardini et al. 1993).
This analysis also shows that Gaussian or skew--positive CDM models suffer
from a lack of large--scale power (regardless of the bias parameter
employed), whereas the skew--negative models have no problem to produce
copious large--scale power. A consistent picture thus emerges: the CDM model
with Gaussian fluctuations cannot account for the properties of clusters in
the Lick map, as measured by a number of independent statistical tests, and,
in order to reconcile CDM with the data, one needs a distribution of
fluctuations with negative initial skewness.

The qualitative agreement between these various statistical descriptions
(number--richness, correlation functions, MST, multifractal, topology,
skewness) demonstrates the usefulness of these methods at quantifying
the large-scale clustering. They all emphasize different aspects of the
clustering pattern -- different descriptors show up certain distinguishing
features of different models -- so they are not all displaying the same
information as contained in the two--point correlation function. The agreement
between the results of our analysis of projected catalogues with similar
analyses of the spatial distribution of matter in these models (Moscardini et
al. 1991; Messina et al. 1992; Coles et al. 1993a; Lucchin et al. 1993) also
shows that using projected
data is still a very worthwhile exercise given the enormous size of the
data sets available. This points is even more remarkable in view of the
recent compilation of extended angular cluster samples, also based on
overdensity criteria, selected from the APM (Dalton et al. 1992) and COSMOS
(Lumsden et al. 1992) galaxy surveys.

In the introduction to this paper, we asked two questions concerning
non--Gaussian primordial fluctuations. The first was whether using
skewed initial data could add sufficient large--scale power to reconcile
the CDM hypothesis with observations. We have only explored a small
(infinitesimal!) subset of all possible non--Gaussian models, but even so
we have found that we can make models that agree much more closely with
the observations than the Gaussian model. We believe therefore that we
have answered this question with a firm `yes'.

The second question that arises is whether it is {\em necessary} to
consider non--Gaussian fluctuations to solve the large--scale structure
problem. Of course, the two--point correlation properties of galaxies
and clusters could be brought into agreement with the data by simply
adding more primordial power to the power--spectrum, either by
changing the background cosmology to have a low value of $\Omega h^{2}$
or by adding a source of large--scale fluctuations such as a Hot
Dark Matter component. Gravitational evolution generates phase correlations
in the non--linear regime in a manner which is coupled to the primordial
power spectrum. It is consequently difficult to distinguish the
effects of extra power from those of intrinsic phase correlations. This
is particularly a problem with angular galaxy catalogues where projection
induces a mixing of length scales which further clouds the issue. All
our simulations incorporate the CDM power spectrum as initial data and
we have not looked at spectra with more primordial power so we cannot say
for certain that non--Gaussian fluctuations are indicated by the observations
regardless of the form of the initial power--spectrum. It is clear, however,
that non--Gaussian CDM models are a viable alternative to the other
solutions to the large--scale structure problem and one should not discard
them until the observations unambiguously rule them out.

\section* {Acknowledgments}
LM acknowledges F. Lucchin, S. Matarrese and A. Messina for the possibility
to use in this paper the results of previous common work on $N$--body
simulations. Moreover we all thank them for useful discussions and fruitful
comments on the manuscript. SB wishes to acknowledge SISSA in Trieste for
its hospitality during several phases of preparation of this work. PC
thanks the Dipartimento di Astronomia at the Universit\`a di Padova for the
hospitality during a visit when some of this work was done. He also
acknowledges support from SERC under the QMW rolling grant GR/H09454. This
work has been partially supported by the Ministero Italiano
dell'Universit\`{a} e della Ricerca Scientifica e Tecnologica and by
Consiglio Nazionale delle Ricerche ({\em Progetto Finalizzato: Sistemi
Informatici e Calcolo Parallelo}). The staff and the management of the
CINECA Computer Centre are warmly acknowledged for their assistance and for
allowing the use of computational facilities.

\newpage
\section*{\center{References}}
\begin{trivlist}
\item[] Abell G.O., 1958, ApJS, 3, 211
\item[] Bahcall N.A., 1988, ARA\&A, 26, 631
\item[] Bahcall N.A., Burgett W.S., 1986, ApJ, 300, L35
\item[] Bahcall N.A., Soneira R.M., 1983, ApJ, 270, 20
\item[] Bahcall N.A., West M.J., 1992, ApJ, 392, 419
\item[] Balian R., Schaeffer R., 1989, A\&A, 226, 373
\item[] Barrow J.D., Bhavsar S.P., Sonoda D.H., 1985, MNRAS, 216, 17
\item[] Bhavsar S.P., Ling E.N., 1988, ApJ, 331, L63
\item[] Borgani S., 1990, A\&A, 240, 223
\item[] Borgani S., 1993, MNRAS, 260, 537
\item[] Borgani S., Jing Y.P., Plionis M., 1992, ApJ, 395, 339 (Paper III)
\item[] Borgani S., Murante G., Provenzale A., Valdarnini R., 1993, Phys.
Rev. A, submitted
\item[] Borgani S., Plionis M., Valdarnini R., 1993, ApJ, 404, 000 (Paper IV)
\item[] Coleman P. H., Pietronero L., 1992, Phys. Rep., 213, 311
\item[] Coles P., 1988, MNRAS, 234, 509
\item[] Coles P., Frenk C.S., 1991, MNRAS, 253, 727
\item[] Coles P., Moscardini L., Lucchin F., Matarrese S., Messina A.,
1993a, MNRAS, submitted
\item[] Coles P., Moscardini L., Plionis M., Lucchin F., Matarrese S.,
Messina A., 1993b, MNRAS, 260, 572
\item[] Collins C.A., Nichol R.C., Lumsden S.L., 1992, MNRAS, 254, 295
\item[] Colombi S., Bouchet F.R., Schaeffer R., 1992, A\&A, 263, 1
\item[] Couchman H.M.P., Carlberg R.G., 1992, ApJ, 389, 453
\item[] Dalton G.B., Efstathiou G., Maddox S.J., Sutherland W.J.,
1992, ApJ, 390, L1
\item[] Davis M., Efstathiou G., Frenk C.S., White S.D.M., 1985, ApJ, 292, 371
\item[] Dekel A., Blumenthal G.R., Primack J.R.,  Olivier S.,
1989, ApJ, 338, L5
\item[] Efstathiou G., Dalton G.B., Sutherland W.J., Maddox S., 1992, MNRAS,
257, 125
\item[]Efstathiou G., Kaiser N., Saunders W., Lawrence A.,
Rowan--Robinson M., Ellis R.S., Frenk C.S., 1990, MNRAS, 247,
10{\small p}
\item[] Grassberger P., Badii R., Politi A., 1988, J. Stat. Phys., 51, 135
\item[] Grassberger P., Procaccia I., 1983, Phys. Rev. Lett., 50, 346
\item[] Groth E.J., Peebles P.J.E., 1977, ApJ, 217, 385
\item[] Hentschel H.G.E., Procaccia I., 1983, Physica, D8, 435
\item[] Jing Y.P., Plionis M., Valdarnini R., 1992, ApJ, 389, 499
\item[] Jing Y.P., Valdarnini R., 1993, ApJ, in press
\item[] Jones B.J.T., Coles P., Mart\'{\i}nez V.J., 1992, MNRAS, 256, 146
\item[] Kaiser N., 1984, ApJ, 284, L9
\item[] Klypin A.A., Kopilov A.I., 1983, Sov. Astr. Lett., 9, 41
\item[] Ling E.N., Frenk C.S., Barrow J.D., 1986, MNRAS, 23, 21{\small p}
\item[] Loveday J., Efstathiou G., Peterson B.A., Maddox S.J., 1992,
ApJ, 400, L43
\item[] Lucchin F., Matarrese S., Messina A., Moscardini L., 1993, in
Proc. of the International School of Physics `E. Fermi' on `Galaxy Formation',
eds. J. Silk and N. Vittorio, in press
\item[] Lumsden S.L., Nichol R.C., Collins C.A., Guzzo L., 1992, MNRAS,
258, 1
\item[] Mandelbrot B.B., 1982, The Fractal Geometry of the Nature,
Freeman \& Co., New York
\item[] Mart\'{\i}nez V.J., Jones B.J.T., Dom\'{\i}nguez--Tenreiro R., van
de Weygaert R., 1990, ApJ, 357, 50
\item[] Matarrese S., Lucchin F., Messina A., Moscardini L., 1991,
MNRAS, 252, 35
\item[] Messina A., Lucchin F., Matarrese S.,  Moscardini L., 1992,
Astroparticle Phys., 1, 99
\item[] Moscardini L., Borgani S., Coles P., Lucchin F., Matarrese S.,
Messina A., Plionis M., 1993, ApJL, submitted.
\item[] Moscardini L., Matarrese S., Lucchin F., Messina A.,
1991, MNRAS, 248, 424
\item[] Olivier S., Blumenthal G.R., Dekel A., Primack J.R.,
 Stanhill D., 1990, ApJ, 356, 1
\item[] Ore O., 1962, Amer. Math. Soc. Colloq. Publ., 38
\item[] Paladin G., Vulpiani A., 1984, Lett. Nuovo Cimento, 41, 82
\item[] Paladin G., Vulpiani A., 1987, Phys. Rep., 156, 147
\item[] Peacock J.A., West M.J., 1992, MNRAS, 259, 494
\item[] Peebles P.J.E., 1980, The Large Scale
 Structure of the Universe, Princeton University Press, Princeton
\item[] Plionis M., Barrow J.D., Frenk C.S., 1991, MNRAS, 249, 662 (Paper I)
\item[] Plionis M., Borgani S., 1991, MNRAS, 254, 306 (Paper II)
\item[] Plionis M., Valdarnini R., 1991, MNRAS, 249, 46
\item[] Plionis M., Valdarnini R., Jing Y.P., 1992, ApJ, 398, 12
\item[] Postman M., Geller M.J., Huchra J.P., 1986, AJ, 91, 1267
\item[] Postman M., Huchra J.P., Geller M.J., 1992, ApJ, 384, 407
\item[] Renyi A., 1970, Probability Theory, North Holland, Amsterdam
\item[] Saunders W. et al.,
 1991, Nat, 349, 32
\item[] Scaramella R., 1992, ApJ, 390, L57
\item[] Scaramella R., Vettolani G., Zamorani G., 1991, ApJ, 376, L1
\item[] Schechter P., 1976, ApJ, 203, 297
\item[] Smoot G.F. et al., 1992, ApJ, 396, L1
\item[] Sutherland W., 1988, MNRAS, 234, 159
\item[] Sutherland W., Efsthatiou G., 1991, MNRAS, 258, 159
\item[] Valdarnini R., Borgani S., Provenzale A., 1992, ApJ, 394, 422
\item[] van de Weygaert R., Jones B.J.T., Mart\'{\i}nez V.J., 1992,
Phys. Lett. A, 169, 145
\item[] White S.D.M., Frenk C.S., Davis M., Efstathiou G., 1987, ApJ,
313, 505
\item[] Yepes G., Dom\'{\i}nguez--Tenreiro R., Couchman H.M.P., 1992,
ApJ, 401, 40
\item[] Zwicky F., Herzog E., Karpowicz M., Koval C.T., 1961--1968,
Catalogue of Galaxies and Clusters of Galaxies. California Institute of
Technology, Pasadena
\end{trivlist}

\newpage

\section*{\center{Figure captions}}
{\bf Figure 1.} The distribution of C25 clusters in the quasi equal area
coordinates defined by eq.(11), for both the `bad' (Fig.1$a$) and the
`good' (Fig.1$b$) models. In both cases, the upper left panel is for the
real clusters. The remarkable difference in the number density of objects
produced by the different models is readily apparent. It is also clear how
clusters can amplify the clustering pattern of the underlying galaxy
distribution (cf. Figure 2 of Coles et al. 1993b).

\vspace{0.2truecm}
\noindent
{\bf Figure 2.} The number of selected clusters $N_{cl}$, the number of
`active' cells belonging to clusters $N_{cells}$ and the cluster
richness $R_1$ (see text) are
plotted from top to bottom, as functions of the cluster identification
threshold $\kappa$ (see eq.[10]). Left and right panels are for `bad' and
`good' models, respectively. Filled circles are for the real clusters, the
open circles for the Gaussian models, the filled squares for
the $\chi^2$ models
and the open squares for the Lognormal models.

\vspace{0.2truecm}
\noindent
{\bf Figure 3.} The 2--point correlation functions, $w(\vartheta)$, for
the `good' models, for each $\kappa$ value. Figure 3$a$, $b$ and $c$ are for
the $G_1$, $\chi^2_n$ and $LN_n$ models, respectively. Filled circles are for
the real data, while open circles are for simulated samples. The dashed and the
dotted straight lines are the power--law best fits to observational and
simulated data, respectively. Error bars represent 1$\sigma$ uncertainties,
obtained through the bootstrap resampling technique.

\vspace{0.2truecm}
\noindent
{\bf Figure 4.} The relation between the correlation amplitude $A$ and the
cluster richness $R_1$. We plot the results for the $G_1$ (open circles),
$\chi^2_n$ (filled squared) and the $LN_n$ models (open squares), as
well as for real clusters (filled circles). Error bars are 1$\sigma$ standard
deviations from the log--log
linear regression of $w(\vartheta)$ in the scale range where the power--law
model is valid (see also Table 2).

\vspace{0.2truecm}
\noindent
{\bf Figure 5.} The frequency distribution of MST branch lengths,
$F(\vartheta)$, for
the C30 and C18 clusters (left and right columns, respectively). Solid
histograms
are for real clusters, while the dashed ones are for the simulated samples.
{}From top to bottom we plot results for $G_1$, $\chi^2_n$ and $LN_n$ models.

\vspace{0.2truecm}
\noindent
{\bf Figure 6.} The same as in Figure 5, but for clusters selected from galaxy
counts in 20$\times 20$ arcmin cells. We only consider the $\kappa =1.8$ case,
since higher thresholds give too few objects at such a larger smoothing scale.

\vspace{0.2truecm}
\noindent
{\bf Figure 7.} The correlation--integral partition function, $Z(q,\vartheta)$
(see eq.[17]), for C36 clusters for $q=0$ (upper panels) and $q=4$
(lower panels).
{}From left to right we report the results for the $G_1$, $\chi^2_n$ and $LN_n$
models. Also plotted is the local dimension, $D_q(\vartheta)$, obtained
according to eq.(\ref{eq:dq1}), from a three--point log--log linear
regression on
the partition function. Filled circles are for PBF clusters (see also Borgani,
Plionis \& Valdarnini 1993), while open squares are for simulations. It
is apparent in the $q=4$ case the difference between the Gaussian model and
the non--Gaussian ones
at reproducing a small--scale flat shape of the local dimension.

\vspace{0.2truecm}
\noindent
{\bf Figure 8.} The same as Figure 7, but for the C25 clusters.

\vspace{0.2truecm}
\noindent
{\bf Figure 9.} The multifractal dimension spectrum, $D_{\tau}$, and the
singularity spectrum, $f(\alpha)$, for the $\chi^2_n$ (left panels) and the
$LN_n$ (right panels) C36 clusters (open circles), as compared to the PBF C36
sample (filled circles). The dimension values are obtained from the slope of
the $W(\tau,p)$ partition function (see eq.[\ref{eq:wp}]) in the $p$ range of
values where a good scaling is observed for both real and simulated data.
Error bars are 1$\sigma$ standard deviations for the log--log linear regression
on the partition function.

\vspace{0.2truecm}
\noindent
{\bf Figure 10.} The variance--skewness relation for the `good' models is
plotted for C30 (left panels) and C18 (right panels) clusters. Filled
triangles are for real clusters, while filled dots are for simulations. The
dashed lines correspond to the hierarchical expression of eq.(\ref{eq:hier}),
with $Q=0.6$.

\newpage
\pagestyle{empty}
\begin{table}[tp]
\centering
\caption[]{Characteristics of the cluster samples. Column 2: number of
selected clusters. Column 3: number of {\em active} cells belonging to
clusters. Columns 4 and 5: cluster richness according to two different
definitions (see text) along with the respective {\em rms} values.}
\tabcolsep 5pt
\begin{tabular}{lrrcccrrcc} \\ \\
 & \multicolumn{4}{c}{Real clusters} & & \multicolumn{4}{c}{ }\\
 ~   & $N_{cl}$ & Cells & $R_1$ & $R_2$ & & & & & \\
 C36 & 285 & 1098 & $5.81\pm 0.72$ & $6.57\pm 1.94$ & & & & & \\
 C30 & 626 & 2720 & $4.69\pm 0.59$ & $5.34\pm 1.67$ & & & & & \\
 C25 &1159 & 5849 & $3.91\pm 0.49$ & $4.49\pm 1.50$ & & & & & \\
 C18 &2685 &20020 & $2.84\pm 0.36$ & $3.34\pm 1.27$ & & & & & \\ \\
 & \multicolumn{4}{c}{$G_{1.5}$} & & \multicolumn{4}{c}{$G_{1}$}\\
 ~   & $N_{cl}$ & Cells & $R_1$ & $R_2$ &
     & $N_{cl}$ & Cells & $R_1$ & $R_2$ \\
 C36 & 45  & 122  & $5.82\pm 0.40$ & $6.15\pm 0.79$ &
     & 131 & 680  & $5.97\pm 0.58$ & $6.58\pm 1.45$ \\
 C30 & 104 & 413  & $4.93\pm 0.40$ & $5.35\pm 0.87$ &
     & 287 & 1575 & $5.00\pm 0.52$ & $5.55\pm 1.36$ \\
 C25 & 306 & 1279 & $4.05\pm 0.37$ & $4.39\pm 0.86$ &
     & 577 & 3524 & $4.17\pm 0.47$ & $4.70\pm 1.27$ \\
 C18 &1569 & 8014 & $2.94\pm 0.25$ & $3.20\pm 0.68$ &
     &1639 &12844 & $3.03\pm 0.35$ & $3.48\pm 1.10$ \\ \\
 & \multicolumn{4}{c}{$\chi^2_p$} & & \multicolumn{4}{c}{$\chi^2_n$}\\
 ~   & $N_{cl}$ & Cells & $R_1$ & $R_2$ &
     & $N_{cl}$ & Cells & $R_1$ & $R_2$ \\
 C36 & 46  & 202  & $6.01\pm 0.61$ & $6.68\pm 1.61$ &
     & 174 & 832  & $5.80\pm 0.52$ & $6.34\pm 1.30$ \\
 C30 & 108 & 523  & $4.98\pm 0.53$ & $5.53\pm 1.44$ &
     & 382 & 2145 & $4.87\pm 0.47$ & $5.37\pm 1.22$ \\
 C25 & 230 & 1244 & $4.13\pm 0.44$ & $4.62\pm 1.28$ &
     & 687 & 4971 & $4.11\pm 0.41$ & $4.61\pm 1.18$ \\
 C18 &1464 & 6875 & $1.92\pm 0.26$ & $3.17\pm 0.76$ &
     &1571 &18684 & $2.94\pm 0.32$ & $3.39\pm 1.10$ \\ \\
 & \multicolumn{4}{c}{$LN_p$} & & \multicolumn{4}{c}{$LN_n$}\\
 ~   & $N_{cl}$ & Cells & $R_1$ & $R_2$ &
     & $N_{cl}$ & Cells & $R_1$ & $R_2$ \\
 C36 & 44  & 242  & $6.08\pm 0.72$ & $6.90\pm 1.98$ &
     & 238 & 1782 & $6.26\pm 0.80$ & $7.13\pm 2.29$ \\
 C30 & 85  & 513  & $5.06\pm 0.65$ & $5.82\pm 1.89$ &
     & 411 & 3767 & $5.27\pm 0.68$ & $6.04\pm 2.08$ \\
 C25 & 190 & 1002 & $4.15\pm 0.58$ & $4.70\pm 1.61$ &
     & 735 & 8011 & $4.34\pm 0.53$ & $4.97\pm 1.81$ \\
 C18 &1658 & 7093 & $2.85\pm 0.25$ & $3.07\pm 0.81$ &
     &1330 &24054 & $3.15\pm 0.36$ & $3.68\pm 1.62$
\end{tabular}
\end{table}


\begin{table}[tp]
\centering
\caption[]{Clusters from 20$\times 20$ arcmin cells. Column 2: number of
selected clusters. Column 3: number of {\em active} cells belonging to
clusters. }
\tabcolsep 15pt
\begin{tabular}{lrrcrr} \\ \\
 & \multicolumn{2}{c}{Real clusters} & & \multicolumn{2}{c}{ } \\
 ~   & $N_{cl}$ & Cells& & & \\
 C25$^{20}$ & 96 & 445 & & &  \\
 C18$^{20}$ &369 &2503 & & &  \\ \\
 & \multicolumn{2}{c}{$G_{1.5}$} & & \multicolumn{2}{c}{$G_{1}$} \\
 ~   & $N_{cl}$ & Cells &  & $N_{cl}$ & Cells \\
 C25$^{20}$ & 26 & 73 & & 86 & 394 \\
 C18$^{20}$ &188 & 886& &311 & 2095 \\ \\
 & \multicolumn{2}{c}{$\chi^2_p$}& & \multicolumn{2}{c}{$\chi^2_n$} \\
 ~   & $N_{cl}$ & Cells &  & $N_{cl}$ & Cells \\
 C25$^{20}$ & 23 & 125& & 126& 625 \\
 C18$^{20}$ &130 & 756& & 333& 3653 \\ \\
 & \multicolumn{2}{c}{$LN_p$}    & & \multicolumn{2}{c}{$LN_n$} \\
 ~   & $N_{cl}$ & Cells &  & $N_{cl}$ & Cells \\
 C25$^{20}$ & 24 & 107& & 146& 1483 \\
 C18$^{20}$ &117 & 621& & 288& 5467
\end{tabular}
\end{table}


\begin{table}[tp]
\centering
\caption[]{Best--fit parameters for the 2--point correlation function,
$w(\vartheta)=A\vartheta^{1-\gamma}$, and the scale range
$[\vartheta_1,\vartheta_2]$ where the power--law is well defined.
The errors are 1$\sigma$ standard deviations.
No results are reported for the C36 clusters of
`bad' models, since they are too few to realize a meaningful correlation
analysis.}
\tabcolsep 7pt
\begin{tabular}{lccccccc} \\ \\
 & \multicolumn{3}{c}{Real clusters} & & \multicolumn{3}{c}{ }\\
 ~   & $A$ & $\gamma$ & $[\vartheta_1,\vartheta_2]$ & & & & \\
 C36 & $1.66\pm 0.25$ & $2.02\pm 0.26$ & 0.6--2.7   & & & & \\
 C30 & $0.96\pm 0.10$ & $2.04\pm 0.09$ & 0.6--8.3   & & & & \\
 C25 & $0.58\pm 0.04$ & $1.98\pm 0.05$ & 0.9--8.3   & & & & \\
 C18 & $0.19\pm 0.01$ & $1.58\pm 0.06$ & 0.9--8.3   & & & & \\ \\
 & \multicolumn{3}{c}{$G_{1.5}$} & & \multicolumn{3}{c}{$G_{1}$}\\
 ~   & $A$ & $\gamma$ & $[\vartheta_1,\vartheta_2]$ &
 ~   & $A$ & $\gamma$ & $[\vartheta_1,\vartheta_2]$ \\
 C36 & -----          & -----          & -----      &
     & $1.26\pm 0.60$ & $2.39\pm 0.43$ & 0.6--8.3 \\
 C30 & $2.14\pm 1.05$ & $2.61\pm 0.40$ & 0.6--8.3   &
     & $0.60\pm 0.13$ & $2.13\pm 0.18$ & 0.6--8.3 \\
 C25 & $1.09\pm 0.28$ & $2.64\pm 0.33$ & 0.6--3.9   &
     & $0.55\pm 0.11$ & $2.53\pm 0.22$ & 0.6--8.3 \\
 C18 & $0.40\pm 0.04$ & $2.77\pm 0.11$ & 0.9--3.9   &
     & $0.24\pm 0.02$ & $2.12\pm 0.10$ & 0.6--3.9 \\ \\
 & \multicolumn{3}{c}{$\chi^2_p$} & & \multicolumn{3}{c}{$\chi^2_n$}\\
 ~   & $A$ & $\gamma$ & $[\vartheta_1,\vartheta_2]$ &
 ~   & $A$ & $\gamma$ & $[\vartheta_1,\vartheta_2]$ \\
 C36 & -----          & -----          & -----       &
     & $2.17\pm 0.34$ & $2.21\pm 0.27$ & 0.6--2.7 \\
 C30 & $2.02\pm 0.65$ & $2.40\pm 0.27$ & 0.6--8.3   &
     & $2.04\pm 0.16$ & $2.21\pm 0.07$ & 0.6--8.3 \\
 C25 & $1.34\pm 0.27$ & $2.22\pm 0.20$ & 0.6--8.3   &
     & $1.24\pm 0.13$ & $2.12\pm 0.09$ & 0.6--8.3 \\
 C18 & $0.48\pm 0.02$ & $2.07\pm 0.06$ & 0.6--3.9   &
     & $0.37\pm 0.02$ & $1.89\pm 0.05$ & 0.6--8.3 \\ \\
 & \multicolumn{3}{c}{$LN_p$} & & \multicolumn{3}{c}{$LN_n$}\\
 ~   & $A$ & $\gamma$ & $[\vartheta_1,\vartheta_2]$ &
 ~   & $A$ & $\gamma$ & $[\vartheta_1,\vartheta_2]$ \\
 C36 & -----          & -----          & -----       &
     & $3.88\pm 0.22$ & $1.95\pm 0.05$ & 0.6--8.3 \\
 C30 & $1.60\pm 0.39$ & $2.18\pm 0.37$ & 0.6--8.3   &
     & $1.74\pm 0.07$ & $1.60\pm 0.04$ & 0.6--8.3 \\
 C25 & $2.39\pm 0.40$ & $2.68\pm 0.21$ & 0.6--8.3   &
     & $1.52\pm 0.11$ & $1.48\pm 0.06$ & 0.6--8.3 \\
 C18 & $0.51\pm 0.03$ & $2.07\pm 0.05$ & 0.6--8.3   &
     & $0.66\pm 0.04$ & $1.53\pm 0.06$ & 0.6--8.3
\end{tabular}
\end{table}


\begin{table}[tp]
\centering
\caption[]{Kolmogorov--Smirnov (KS) test for the MST branch--length frequency
distribution, $F(\vartheta)$ (see text). Column 2: the KS statistics,
$\bar D$, measuring the maximum difference between the $F(\vartheta)$
cumulative distributions for real and simulated cluster samples. Column 3:
probability to measure a larger difference from a statistically equivalent
realization; it is the
significance level of the difference; smaller values correspond to more
significative differences. We also present results for the C18$^{20}$
samples, which contains clusters selected from smoothed counts in $20\times
20$ arcmin cells (see also Table 2).}
\tabcolsep 15pt
\begin{tabular}{lccccc} \\ \\
 & \multicolumn{2}{c}{$G_{1.5}$} & & \multicolumn{2}{c}{$G_{1}$} \\
 ~   & $\bar D$  & $P(D>\bar D)$ & & $\bar D$ & $P(D>\bar D)$ \\
 C36 & 0.10 & 0.86               & &  0.15 & 0.04 \\
 C30 & 0.18 & $6.2~ 10^{-3}$     & & 0.12 & $7.5~ 10^{-3}$ \\
 C25 & 0.10 & 0.02               & & 0.10 & $5.2~ 10^{-4}$ \\
 C18 & 0.07 & $7.3~ 10^{-5}$     & & 0.06 & $2.6~ 10^{-3}$ \\
 C18$^{20}$ & 0.08 & 0.44        & & 0.05 & 0.88           \\ \\
 & \multicolumn{2}{c}{$\chi^2_p$}& & \multicolumn{2}{c}{$\chi^2_n$} \\
 ~   & $\bar D$  & $P(D>\bar D)$ & & $\bar D$ & $P(D>\bar D)$ \\
 C36 & 0.12 & 0.62               & & 0.12 & 0.09 \\
 C30 & 0.11 & 0.17               & & 0.09 & 0.04 \\
 C25 & 0.12 & $8.3~ 10^{-3}$     & & 0.11 & $1.1~ 10^{-4}$ \\
 C18 & 0.08 & $8.3~ 10^{-6}$     & & 0.09 & $1.4~ 10^{-6}$ \\
 C18$^{20}$ & 0.10 & 0.36        & & 0.03 & 0.99 \\ \\
 & \multicolumn{2}{c}{$LN_p$}    & & \multicolumn{2}{c}{$LN_n$} \\
 ~   & $\bar D$  & $P(D>\bar D)$ & & $\bar D$ & $P(D>\bar D)$ \\
 C36 & 0.22 & 0.09               & & 0.12 & 0.08 \\
 C30 & 0.14 & 0.11               & & 0.05 & 0.67 \\
 C25 & 0.11 & 0.05               & & 0.12 & $3.4~ 10^{-5}$ \\
 C18 & 0.08 & $3.2~ 10^{-5}$     & & 0.09 & $1.6~ 10^{-6}$ \\
 C18$^{20}$ & 0.10 & 0.45        & & 0.06 & 0.75
\end{tabular}
\end{table}


\begin{table}[tp]
\centering
\caption[]{The coefficient $Q$ in the variance--skewness relation (see
eq.[25]), for real data and `good' models. The errors are
1$\sigma$ standard deviations in the $\log \Gamma$--$\log \Sigma^2$ linear
regression.}
\tabcolsep 15pt
\begin{tabular}{lccc} \\ \\
 & Real clusters & & $G_{1}$ \\
 C36 & $0.54 \pm 0.28$ & & $0.53\pm 0.27$ \\
 C30 & $0.71 \pm 0.37$ & & $0.60\pm 0.32$ \\
 C25 & $0.65 \pm 0.36$ & & $0.63\pm 0.35$ \\
 C18 & $0.83 \pm 0.50$ & & $0.70\pm 0.42$ \\ \\
 & $\chi^2_n$    & & $LN_n$   \\
 C36 & $0.58 \pm 0.30$ & & $0.59\pm 0.31$ \\
 C30 & $0.65 \pm 0.33$ & & $0.57\pm 0.31$ \\
 C25 & $0.59 \pm 0.30$ & & $0.62\pm 0.32$ \\
 C18 & $0.66 \pm 0.38$ & & $0.68\pm 0.40$
\end{tabular}
\end{table}

\end{document}